\documentclass[reprint,amsmath,assume,aps]{revtex4-2}
\usepackage{graphicx}
\usepackage{dcolumn}
\usepackage{bm}
\usepackage{placeins} 
\usepackage{xspace}
\usepackage{xcolor} 
\usepackage{lineno}
\usepackage{hyperref}

\begin{document}
\newcommand{\evtlib}{\texttt{evtLib}\xspace} 

\preprint{APS/123-QED}

\title{Monte Carlo Simulation Development and Implementation of the GiBUU Model for Neutrino Experiments}

\author{L.~Aliaga}
\affiliation{University of Texas at Arlington}
\author{R.~Castillo~Fern\'andez}
\affiliation{University of Texas at Arlington}
\author{J.~Gustafson}
\affiliation{University of Texas at Arlington}
\author{D.~Quinn}
\affiliation{University of Texas at Arlington}
\author{S.~Yadav}
\affiliation{University of Texas at Arlington}

\date{\today}

\begin{abstract}
This paper introduces a Monte Carlo simulation generated with the GiBUU model for neutrino experiments. The simulation generates realistic neutrino event samples, contributing to the prediction and interpretation of experimental outcomes. The results showcase the performance of the GiBUU-based simulation framework, emphasizing its fidelity to the original GiBUU cross-section model. This first implementation enables future work on developing the infrastructure to propagate systematic uncertainties. These contributions enhance the precision of experimental predictions and provide a platform for further exploration in future studies.
\end{abstract}

\maketitle

\section{Introduction}
\label{sec:Intro}

Neutrinos, elusive and ghostly particles, have captivated the attention of physicists for decades. Neutrinos possess intriguing properties, such as their weak interactions and tiny masses. These characteristics make neutrinos fascinating subjects of study, with profound implications for our understanding of the fundamental laws of nature. Neutrino physics is essential in particle physics, offering unique insights into the fundamental forces and particles that govern the universe. 

Neutrino masses are demonstrated because they oscillate between flavors (electron, muon, and tau) as they propagate through space, and this phenomenon provides unequivocal evidence for physics beyond the Standard Model~\cite{bib:osctheory1, bib:osctheory2}. Neutrino oscillation experiments have led to groundbreaking discoveries, challenging our understanding of neutrino properties and paving the way for further investigations beyond the known realm.

In particular, accelerator-based neutrino experiments have significantly advanced our understanding of neutrinos and their properties. These experiments employ powerful particle accelerators to produce intense neutrino beams, allowing for detailed investigations of neutrino interactions, oscillations, and their potential implications for other physics beyond the Standard Model. Understanding neutrino oscillations is paramount for advancing our knowledge of the lepton sector and exploring new physics; for example, deviations from the observed oscillation patterns could indicate the presence of new neutrino flavors or interactions that go beyond the established framework of the Standard Model. Moreover, these experiments provide a unique avenue for probing the Universe's asymmetry between matter and antimatter.

In this context, Monte Carlo simulations are essential for designing, analyzing, and interpreting experimental data. These simulations provide a means to model and predict neutrino interactions with target materials, detector responses, and backgrounds, thus enabling precise estimates of the expected event rates to extract the desired physics parameters, including corrections due to limited efficiencies and backgrounds. The importance of accurate models in Monte Carlo simulations stems from their significant impact on result extraction and the subsequent interpretation of physics outcomes. The relevance of our Monte Carlo generator extends beyond the accurate modeling of neutrino-nucleus cross sections; it enables precise predictions of neutrino event rates, energy spectra, and particle kinematic distributions. These predictions are crucial for the interpretation of oscillation measurements and the search for new physics phenomena. 

This paper presents a Monte Carlo event generator based on the GiBUU (Giessen Boltzmann-Uehling-Uhlenbeck) model~\cite{bib:GiBUU1,bib:GiBUU2} and specifically tailored for neutrino experiments. The output of GiBUU is a weighted event simulation, which adds difficulty to its implementation as an event simulation within a detector geometry. Within this work, the GiBUU model, renowned for its success in describing hadron/electron/photon/neutrino-nucleus interactions in various target materials, is converted into an event-by-event predictor, or in other words, an unweighted event generator. By incorporating the GiBUU model into a Monte Carlo simulation framework, we aim to provide a comprehensive tool for simulating and analyzing neutrino data events. An important aspect of neutrino oscillations is reconstructing the neutrino energy, mostly relying on the final state particles observed from the neutrino interaction. The model approach adopted by GiBUU with its nuclear transport model, different from GENIE and other neutrino event generators, can offer important insights to understand those final state particles and improve our accuracy to assess uncertainties in corrections due to inefficiencies and backgrounds. GiBUU cannot account for detector geometry, so it was unsuitable to be implemented as an event generator for experiments.

The simulation leverages the GENIE~\cite{bib:GENIE1,bib:GENIE2} geometry and flux drivers and produces the output event formats compatible with those from GENIE. By adopting this approach, we can run experiment-specific software to reconstruct the generated events in the same manner as we would with GENIE-generated events, eliminating the burden of developing additional software at each experiment to read GiBUU Monte Carlo outputs. The generated events can be efficiently propagated to simulate detector responses using experiment-specific software. This process accurately reproduces observed experimental signatures with the same software setup used for neutrinos generated by the GENIE Monte Carlo. Notably, the developed method also applies to other event generators, which could be especially beneficial for those producing weighted events, a common feature in many Beyond the Standard Model simulations.

This implementation has been proved into the Neutrino Event Generation stage of LArSoft (Liquid Argon Software~\cite{bib:Snider:2017wjd}) $v09\_53\_02$, and it has been tested using the SBND~\cite{bib:SBNDarxiv} and ICARUS~\cite{bib:ICARUSOpsarxiv} detector geometries. Both geometries are traversed by an enhanced $\nu_{\mu}$ beam from the BNB (Booster Neutrino Beam) beamline at Fermilab~\cite{bib:BNBflux}. The BNB flux peaks at 0.6-0.8 GeV in these experiments. The distributions presented in this report were generated using a simulated box geometry with dimensions of 2 m on each side and filled with Liquid Argon, located 110 m from the BNB target, referred to as LArBox throughout the paper. 

Section~\ref{sec:GiBUU} offers a brief description of the GiBUU model that is relevant for neutrino-nucleus scattering in the context of neutrino oscillations. Section~\ref{sec:MCFrame} provides an overview and discussion of how the Monte Carlo event generator is produced. A brief discussion of the following steps towards calculating and implementing the systematic uncertainties associated with the GiBUU model is presented in Section~\ref{sec:Systematics}, and the results of the work, including comparisons to GENIE, and final remarks are presented in Section~\ref{sec:Results}.

\section{ The GiBUU Model and Its Relevance to Neutrino Interactions}
\label{sec:GiBUU}

The GiBUU~\cite{bib:GiBUU1,bib:GiBUU2} model serves as a comprehensive theoretical framework designed to study interactions of hadrons, electrons, and neutrinos within various nuclear systems. Grounded in kinetic theory and effective quantum field theory principles, GiBUU incorporates nuclear transport theory and mean-field dynamics concepts to describe the evolution of hadronic systems by considering their interactions through scattering, decay, and production processes. The GiBUU model has been extensively used in a range of research areas, including nuclear reactions, heavy-ion collisions, and neutrino-nucleus interactions, the latter being the focus of this work. Its versatility and accuracy have made it valuable in interpreting experimental data and understanding the underlying dynamics of nuclear systems. 

An essential feature of the GiBUU model is its remarkable accuracy in replicating various experimental datasets without the need for tuning to neutrino scattering data, in which several assumptions of the initial neutrino would have been to be made. To develop some of the model parametrizations, the GiBUU model is informed by comparisons to electron and hadron experimental scattering data, where precise knowledge of the incoming particle energy is available~\cite{leitner2009hadronictransportapproachneutrino}. GiBUU has been extensively tested and validated against various measurements, demonstrating its stability and robustness in describing neutrino-nucleus interactions, including argon targets~\cite{bib:uBooNENuMuCCInc}. This unique feature of the GiBUU model highlights its inherent ability to capture the intricacies of neutrino interactions without relying on specific adjustments or tuning procedures to match neutrino experimental results. Therefore, using the GiBUU model to calculate efficiency, background description, and other model-dependent measurement corrections can provide important insights to understand uncertainties in the physics models and prevent shortcomings. Understanding how the uncertainties are associated with the model, without relying on neutrino data tuning, enables a more robust handling of related uncertainties. By leveraging the inherent accuracy and stability of the GiBUU model, the systematic uncertainties associated with neutrino experiments can be effectively addressed and can help us to advance the efforts to minimize those uncertainties, enhancing the reliability and precision of the results obtained.

GiBUU treats the neutrino-nucleus interaction using weak charged current (CC) and neutral current (NC) interaction processes. It considers the quasi-elastic scattering (QE), resonant and non-resonant production mechanisms, deep inelastic scattering (DIS), and meson-exchange current processes that play significant roles in neutrino interactions. The nuclear ground model uses a Woods-Saxon density distribution while ensuring a constant Fermi-energy. Some of the neutrino-nucleus interaction mechanisms, such as 1-pion background and 2-pion background, are modeled using MAID~\cite{Drechsel:1998hk} predictions, while PYTHIA~\cite{Sjostrand:2007gs} predictions are applied for DIS. Quasi-elastic and resonant processes are treated simultaneously when calculating lepton-nucleon interaction cross-sections for a single-hadron final state as quasi-elastic scattering and resonance excitation. Other processes, such as lepton-nucleon interaction cross sections for a two-hadron final state, contribute to $N\pi$ backgrounds. Meson exchange current interactions are implemented to CC and NC events using the same model from Bosted and Christy~\cite{bosted2012empiricalfitelectronnucleusscattering}. For a more detailed description of the interaction models, see~\cite{Gallmeister_2016}.

The GiBUU model incorporates nuclear effects through a comprehensive treatment of Final State Interactions (FSI) with nucleon-level interactions. FSI accounts for the intricate interplay between the produced hadrons and the encompassing nuclear medium, including rescattering absorption and re-emission processes. To meticulously capture these interactions with nuclei, the model integrates nuclear structure information, maintains a coherent nuclear potential description, and employs the transport model (the BUU equation). This approach ensures a precise treatment of nuclear effects and elucidates the nuanced influence of a nucleus on these interactions. In GiBUU, this facet proves indispensable for the comprehension of hadronic observables arising from the presence of nuclear matter. The FSI model in GiBUU is more sophisticated than in other event generators, and systematic uncertainties due to FSI may differ from the approach followed by other event generators. The work to understand these uncertainties and how to implement them is ongoing. This aspect is crucial for neutrino experiments analyzing exclusive neutrino scattering topologies, in which the description of each produced particle leaving the nucleus plays an essential role in neutrino energy reconstruction. 

The GiBUU model undergoes rigorous validation against diverse neutrino experimental datasets, encompassing charged-current and neutral-current interactions~\cite{bib:GiBUU3,bib:GiBUU4}. Though tensions with some datasets have been noted—through prior data corrections using alternative Monte Carlo methods—GiBUU's overarching performance remains notably robust. Its capability to faithfully replicate experimental observables, spanning cross-sections, energy spectra, and angular distributions, attests to its reliability and accuracy in portraying the intricate dynamics inherent to neutrino interactions. While recognizing localized disparities, the overarching proficiency of GiBUU positions it as a formidable tool for comprehensive analyses in neutrino physics.

\section{\label{sec:MCFrame}Monte Carlo Simulation Framework for Neutrino Experiments}

One of the most common neutrino Monte Carlo event generators, GENIE~\cite{bib:GENIE1,bib:GENIE2}, has emerged as an essential tool for simulating neutrino interactions in various experiments, and in particular for liquid argon experiments at Fermilab, such as  MicroBooNE~\cite{bib:uBooNESterile2022}, ICARUS~\cite{bib:ICARUSOpsarxiv}, SBND~\cite{bib:SBNDarxiv} and DUNE~\cite{bib:DUNETDR} in which coordinated efforts for development, tuning, and uncertainty estimation of the GENIE generator are already formed. By generating simulated events that mimic real neutrino interactions, GENIE facilitates interpreting data obtained from neutrino experiments and aids in developing and optimizing experimental designs. The widespread utilization of the GENIE event generator in numerous neutrino experiments establishes its critical role. Consequently, it is unsurprising that any subsequent software infrastructure development encompassing detector simulation and event reconstruction is inherently compatible with the output format of GENIE for many experiments and, in particular, for USA-based experiments. Hence, within this work, the simulation of neutrino events using the GIBUU model, considering a neutrino beam and detector geometry (LArBox), utilizes the GENIE flux and geometry drivers. Furthermore, the resulting generated GiBUU events are formatted following GENIE's format to ensure seamless integration and compatibility.

Our simulation consists of two distinct steps. 

Firstly, a significant number of GiBUU events are simulated utilizing the original GiBUU model framework that the Giessen group developed. We produced $1.3 \times 10^{10}$ events aiming to generate extensive libraries of neutrino events. This approach ensures event uniqueness and minimizes repetitions for each neutrino flavor and interaction type based on a given neutrino flux. 

The original GiBUU simulation provides cross-sections, including the squares of the interaction amplitudes, some carrying negative weights, signifying their contribution as destructive interference terms in the overall cross-section calculation. These negative weights arise from non-resonant pion background processes, particularly relevant at higher energies such as those expected at the DUNE experiment. For example, for pion production, the sum of the square of the background amplitude and the interference term of resonance and background amplitudes can be negative. This contribution is the event type that describes the $1\pi^+$ background plus interference. 

Each entry of the libraries stores the kinematic information of the incoming neutrino and produced particle kinematics, its GiBUU weight, and the event generator reaction code, which denotes the initial neutrino-nucleus interaction mode. Two libraries are created; one contains events with positive weights and the other with negative weights. In the latter case, the negative weights are converted to positive by taking their absolute values. 

\begin{figure}[h]
    \centering
    \includegraphics[width=0.45\textwidth]{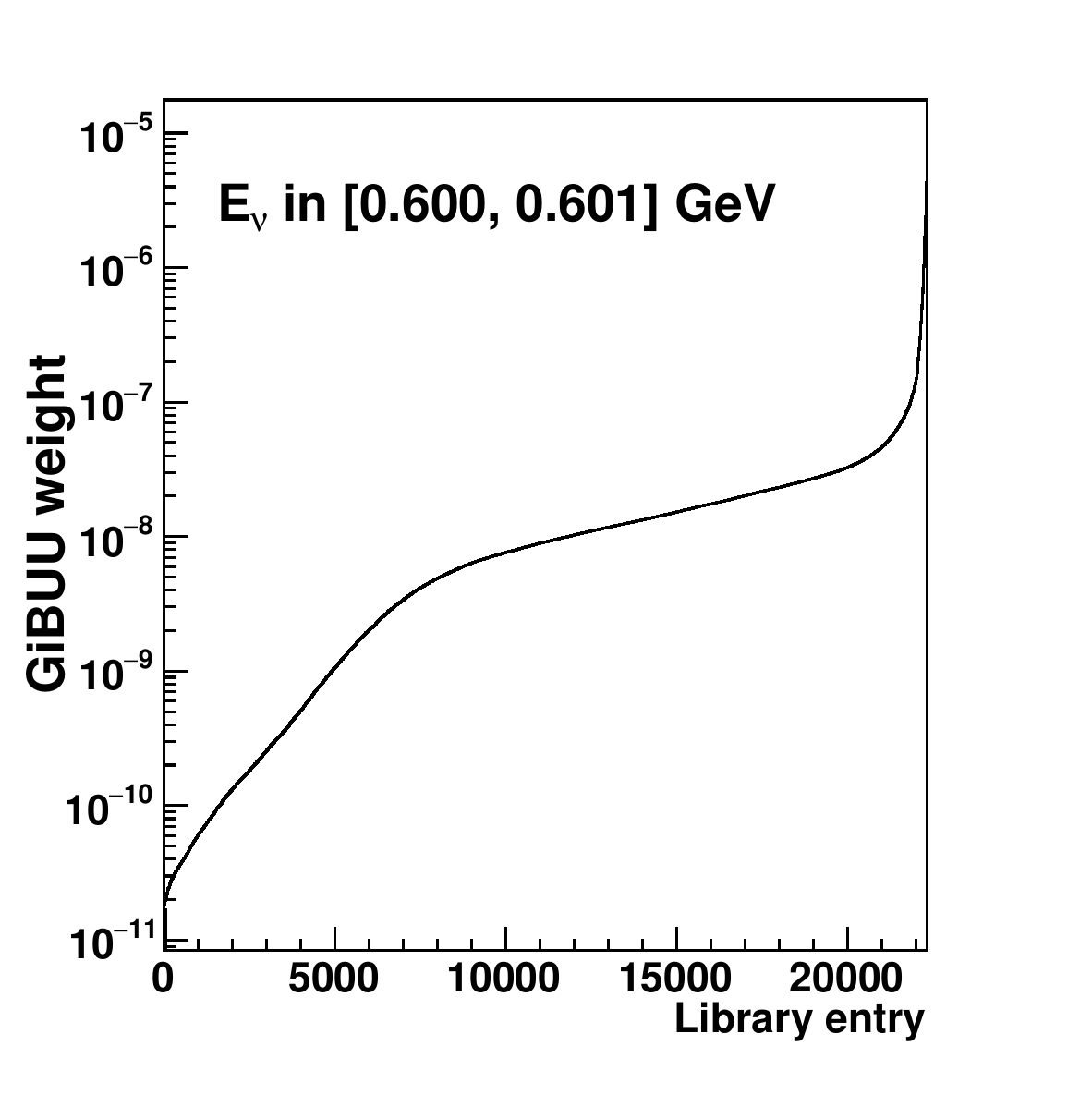}
    \caption{Distribution of weights generated in the GiBUU library for this study, illustrating muon neutrino charged-current interactions at a neutrino energy range of $0.6-0.601$ GeV.}
    \label{fig:gwgts}
\end{figure}

Secondly, two independent productions are made, one per library set, using an acceptance-rejection method, event by event, to select the input from the library. The weight distribution from narrow neutrino energy in the library is used as probability distribution function input to the method, like the sample shown in Fig.~\ref{fig:gwgts}. This visualization also provides insights into the scale of the generated libraries and the corresponding weight yields within a narrow energy range. Our strategy is to run any high-level analysis with the two productions and to calculate the desired final kinematic distribution by subtracting them. 

\begin{figure}[!htbp]
    \centering
    \includegraphics[width=0.45\textwidth]{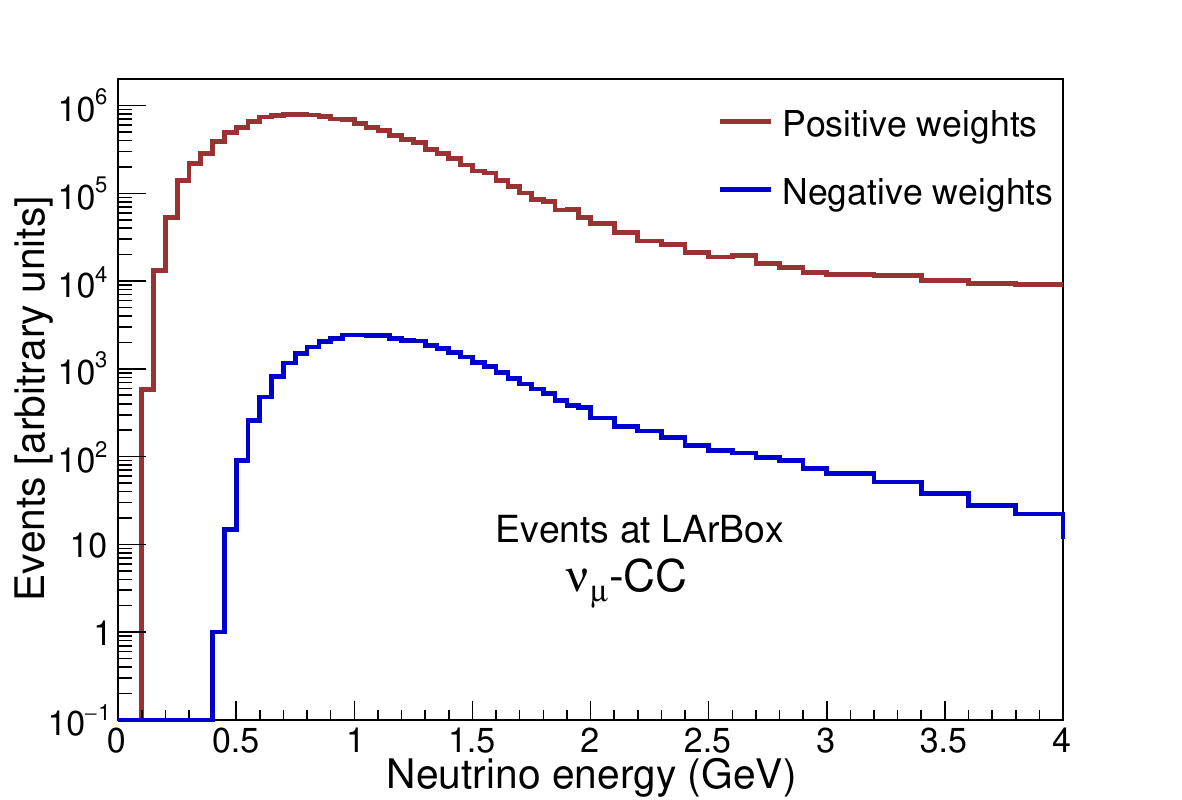}
    \caption{Neutrino energy distribution in the LArBox geometry for muon neutrino charged-current interactions, showing both "negative" and "positive" weight contributions.}
    \label{fig:allnusPosNeg}
\end{figure}

Furthermore, in addition to building a comprehensive statistical library of GiBUU events, we apply interaction vertex rotation techniques to avoid event duplication. Figure \ref{fig:allnusPosNeg} shows the neutrino energy distributions for muon-neutrino charged-current ($\nu_{\mu} CC$) events for both productions in the LArBox geometry. The "negative weights" curve is significantly lower than the "positive weights" at these energies.

Our code is a customized version of the GENIE \evtlib~\cite{bib:GENIE2}, which diverges from the original in several aspects. Notably, our customized \evtlib can handle weighted events and crucially preserves the event identity, retaining the information related to the event generator reaction code (QE, resonant, 2p2h,..), run, and event ID as a unique identifier per event, as well as neutrino energy, weight, and an array of final state particles with their kinematics. Subsequently, these events are incorporated into the LArSoft framework as an external library. During this stage, the flux and detector geometry drivers developed within the GENIE framework are employed to place the vertices for the pre-calculated cross sections from the GiBUU libraries. The same methodology could be used for other cross-section models not yet available as event generators, such as some Beyond the Standard Model theory simulations. 

Our simulation study employed the described technique to simulate GiBUU events within the LArBox geometry, utilizing the BNB (Booster Neutrino Beam) neutrino beam~\cite{bib:BNBflux} prediction. For the studies presented in this paper, we used the GHEP event format. However, it is important to note that other event formats available in GENIE can be used, such as GST, rootracker, and other future event generator formats, such as NuHEPMC, that may be included in GENIE. This facilitates our software implementation in any experiment already using GENIE.

\begin{figure}[!htbp]
    \centering
    \includegraphics[width=0.45\textwidth]{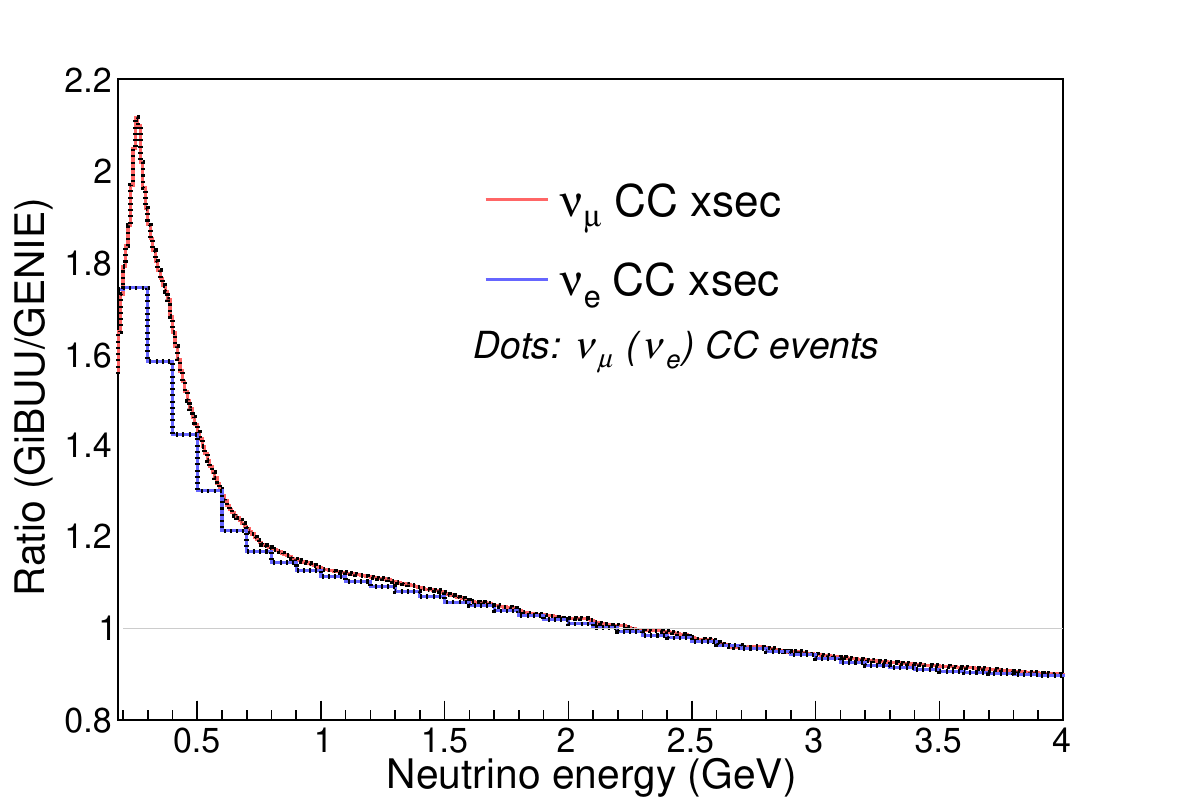}
    \caption{Validation plot comparing the ratios of inclusive charged-current cross sections (GiBUU over GENIE, shown with solid lines) compared with ratios of generated events (shown with dotted lines) using equivalent neutrino flux exposure.}
    \label{fig:ratioxsecevts}
\end{figure}

Following this, we conducted several validation procedures on the implementation, including verifying the accuracy of the event rate calculation. Figure \ref{fig:ratioxsecevts} shows the ratios of inclusive charged-current cross sections (GiBUU over GENIE) as a function of neutrino energy, compared with ratios of generated events with equivalent neutrino flux exposure, demonstrating that both ratios are identical.

\begin{figure}[!htbp]
    \centering
    \includegraphics[width=0.45\textwidth]{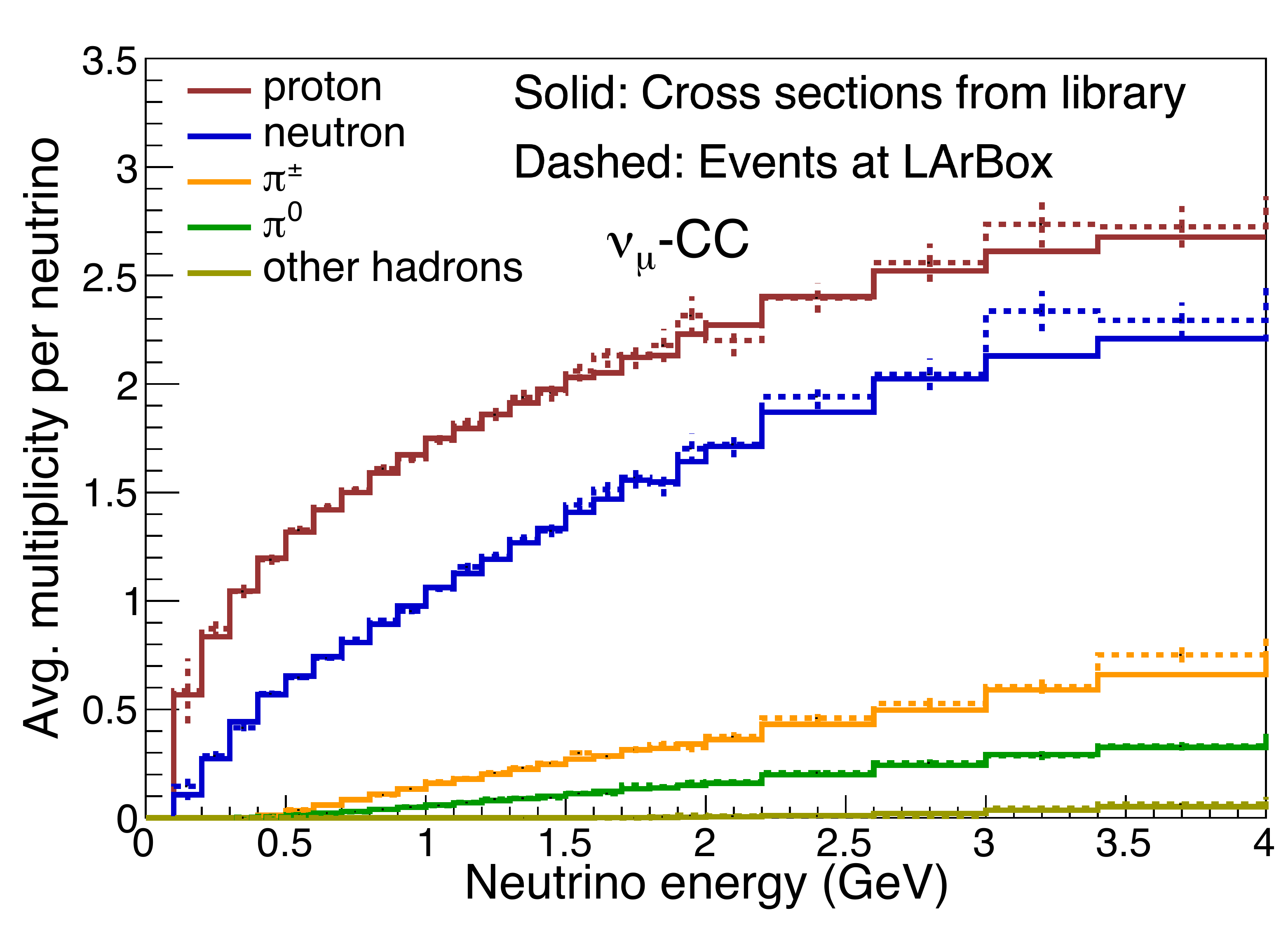}
    \includegraphics[width=0.45\textwidth]{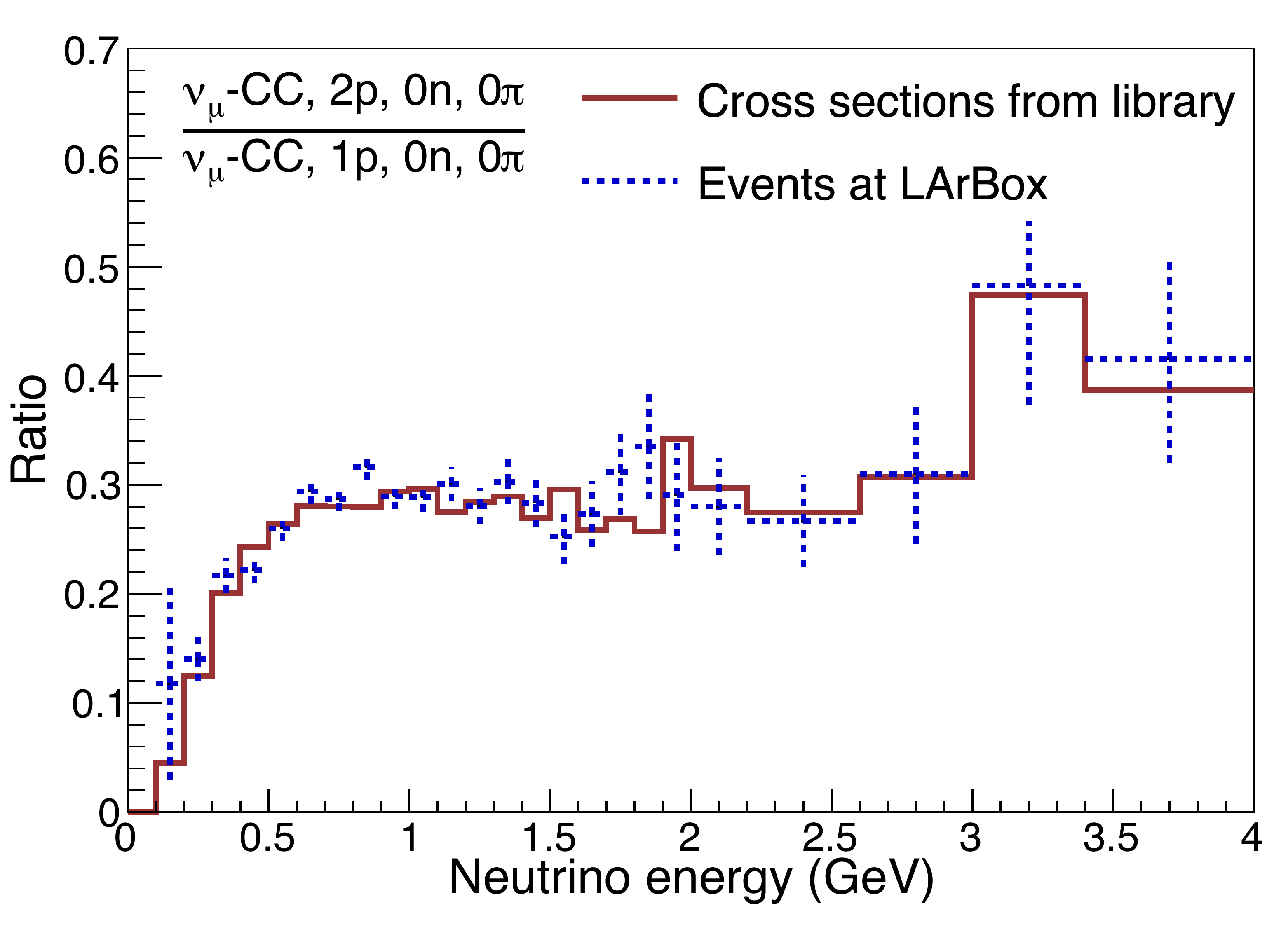}
    \caption{Validation plots comparing the GiBUU cross-section library with GiBUU-generated events in the LArBox geometry as a function of neutrino energy. Top plot: Average hadron multiplicity per $\nu_{\mu}$ CC interaction. Bottom plot: Ratio of a selection enhancing MEC (one muon, two protons, zero pions) to a selection enhancing QE interactions (one muon, one proton, zero pions).}
    \label{fig:hadval}
\end{figure}

Additional validations were performed to verify the authenticity of the produced events by comparing them to the original GiBUU cross-section library, ensuring their conformity to the original GiBUU model. Figure \ref{fig:hadval} shows two of these studies looking at $\nu_{\mu} CC$ events with respect to the neutrino energy: the average hadron multiplicity and the ratio between events with a MEC-enhanced selection (one muon, two protons, and zero pions in the final state) and a QE-enhanced selection (one muon, one proton, and zero pions in the final state). Within the different tests, we studied the topological description of the simulated neutrino events and the contributing neutrino-nucleus interaction types. The interaction types, such as QE, 2p2h, resonant, and DIS, provide essential information to understand the details of the modeling and comparisons to other generators. 

Besides statistical fluctuations, especially at larger energies (where the BNB flux at LArBox is negligible) and minor differences due to the geometrical effects, the validation was successfully completed. The validation tests confirmed the GiBUU model's transformation into a Monte Carlo event generator incorporating the appropriate flux and geometry settings.

However, a topological event description, such as classifying events according to their final state particles, is also required to better understand contributions from final-state interactions (FSI). Comparing the final-state particle contribution per event to measured data is the most effective approach, as only these final-state particles can be detected in experiments. Furthermore, different neutrino-nucleus interactions and FSI can lead to similar observed topologies. It is also important to note that in GiBUU, final state protons below 9.38 MeV in kinetic energy in argon interactions should not be considered final state particles as they may still be bound to the nucleus. In the studies presented in Section \ref{sec:Results}, we incorporate a minimum kinetic energy threshold at each hadron to describe each topology to anticipate the detectability capabilities of current LArTPCs and avoid bound protons with low kinetic energy.

\section{Ongoing Developments in Systematic Uncertainty Estimation for GiBUU}
\label{sec:Systematics}

As neutrino physics advances and experimental data becomes increasingly precise, the need to reduce systematic uncertainties in theoretical models like GiBUU becomes more pressing. Collaborative efforts should exist to refine these model aspects, ultimately enhancing our ability to extract meaningful physics results.

Like any model, the GiBUU framework is subject to systematic uncertainties that can impact the precision of its predictions. After converting GiBUU into a Monte Carlo, the next step is to assess these systematic uncertainties and develop the software framework necessary to propagate those uncertainties. This section aims to delve into the areas of systematic uncertainty estimation that need to be designed for GiBUU. The inherent uncertainties associated with nuclear physics inputs can introduce systematic biases in the simulation results. Understanding and quantifying these uncertainties is crucial for robust data analysis and extracting meaningful physics results. For a deeper discussion on the models we will refer to~\cite{Gallmeister_2016}.

We observe that various sources contribute to systematic uncertainties in the GiBUU model: 
\paragraph{Nuclear Effects and Medium Modifications} One of the potentially primary sources of systematic uncertainty in the GiBUU model arises from nuclear effects and medium modifications. While the model incorporates nuclear structure information and accounts for FSI, there is ongoing research to improve our understanding of these phenomena:

\begin{itemize}
    \item Nuclear density and structure: GiBUU relies on nuclear density distributions and form factors to describe the presence of nuclei. Variations in these parameters can introduce uncertainties in the simulation results. 
    \item Medium Modifications: GiBUU considers modifying hadron properties within the nuclear medium. This feature includes hadron masses and width changes due to interactions with surrounding nucleons. A deeper exploration of the dynamics of these medium modifications is necessary to refine uncertainty estimates.
\end{itemize}

\paragraph{Resonance Contributions} GiBUU incorporates resonance dynamics, but estimations on the uncertainties associated must be addressed:
\begin{itemize}
    \item Resonance Widths: The precise determination of resonance widths and their energy dependence remains challenging. Variations in the assumed resonance widths can lead to uncertainties in resonance production and decay contributions.
    \item Coupling Constants: The coupling constants governing resonance interactions are parameters with associated uncertainties.
\end{itemize}

\paragraph{Neutrino-Nucleon Cross Sections} Accurate cross sections are crucial for reducing systematic uncertainties, but challenges persist:

\begin{itemize}
    \item Energy Dependence: Cross sections depend on energy, and uncertainties in these dependencies can propagate into the predictions. 
    \item Nuclear Environment: The cross sections within the nuclear medium differ from those in free space due to nuclear correlations and uncertainties in these models, especially in neutrino scattering, where constraining these uncertainties using neutrino data is more challenging.
\end{itemize}

Systematic uncertainties arising from non-resonant pion backgrounds and other reaction processes, such as two-particle two-hole interactions and those specific to hadronization modeling, are not explicitly addressed here but will need to be developed to assess GiBUU's systematic effects. Conversely, processes modeled by GiBUU, like the non-resonant pion background, could provide valuable insights into uncertainties resulting from the absence of these interactions in other neutrino event generators, such as GENIE, particularly in the context of DUNE.

Comparisons between GiBUU simulations and experimental data are crucial for assessing systematic uncertainties and validating the model. We can identify some of the associated uncertainties by comparing simulated event observables to data from electron and hadron scattering experiments. However, comparisons to neutrino scattering data are also critical. 

The development of the software infrastructure for GiBUU systematics and its evaluation is an ongoing effort that will be described in a future publication. Currently, this work uses data to evaluate uncertainties and generate different event libraries with these variations. 

A package we are implementing, GiBUU-Systematics, will handle all systematics and will interface with tools used in neutrino experiments. The main difference with other current neutrino generator reweighting tools is that GiBUU-Systematics produces systematic variations simultaneously with the central value library without interfering with each other, storing them in a separate library. This allows experiments to generate both central values and systematic variations per neutrino event, making them immediately available to analyzers. Producing systematic variations beforehand, rather than reweighting afterward, is a novel approach we believe offers benefits without compromising physics accuracy. 

To assess uncertainties within GiBUU-Systematics, we employ a three-step evaluation process, taking advantage that GiBUU is not tuned to neutrino data. First, we compare with hadron scattering data, which is sensitive to strong interactions and uses highly accurate beams. The second step involves comparing with electron scattering data, which is sensitive to electroweak interactions and offers similar beam accuracy. These first two steps help capture uncertainties that affect the final state of the interaction and some accessible electroweak parameters.

In the third step, we compare with neutrino scattering data to capture uncertainties not visible in hadron and electron scattering. Neutrino interactions in nuclei are sensitive to weak interactions, though particles produced are also subject to strong interactions, impacting the event’s final state. Since factorization of effects in neutrino data is  challenging and may be impossible, and beam uncertainties are significant due to neutrino production mechanisms, this comparison step is the final one.

\section{\label{sec:Results}Results and Discussion}

\begin{figure}[!htbp]
    \centering
    \includegraphics[width=0.45\textwidth]{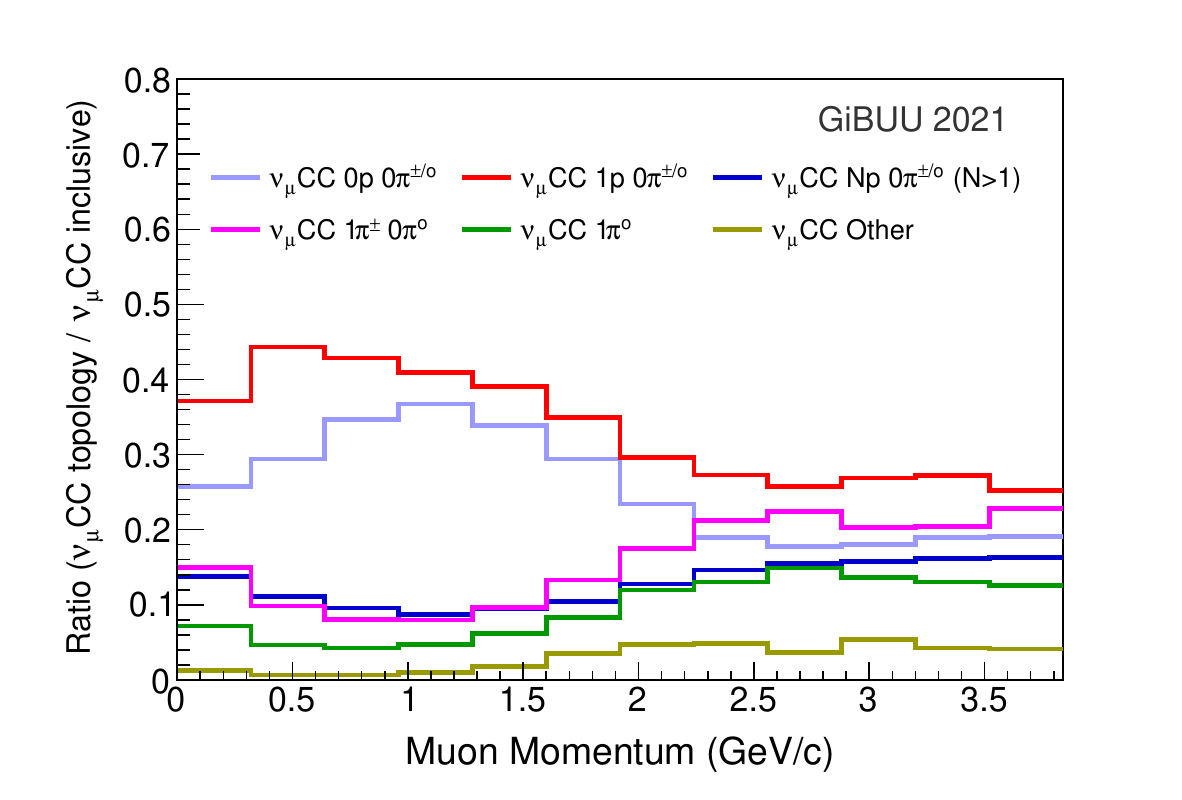}
    \includegraphics[width=0.45\textwidth]{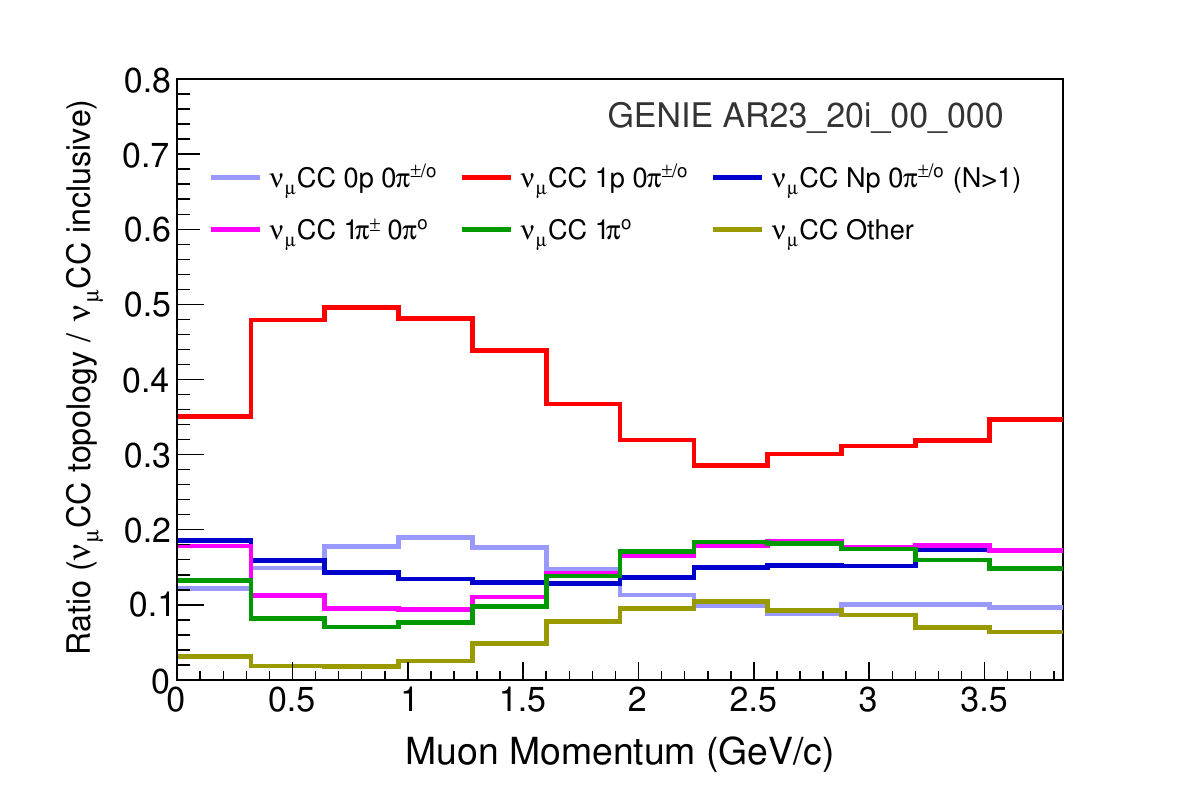}
    \includegraphics[width=0.45\textwidth]{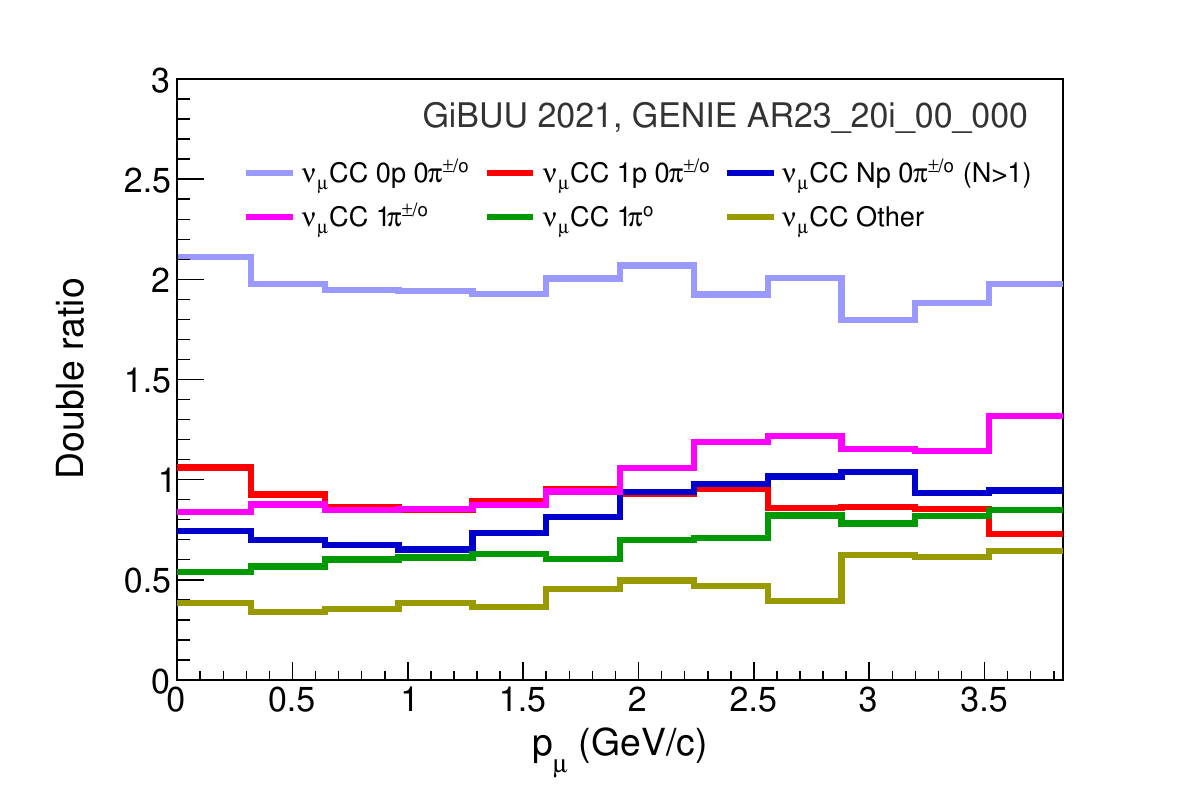}
    \caption{Ratios of each exclusive topology to the inclusive charged-current muon neutrino prediction as a function of muon momentum, comparing GiBUU and GENIE in the LArBox geometry with the BNB flux. The top and middle plots show the ratios for GiBUU and GENIE, respectively. The bottom plot displays the double ratio, calculated as the ratio between the top and middle plots. Proton and pion kinetic energy thresholds are set at 50 MeV to define the topologies.}
    \label{fig:comptGenieGiBUU_mom}
\end{figure}

\begin{figure}[!htbp]
    \centering
    \includegraphics[width=0.45\textwidth]{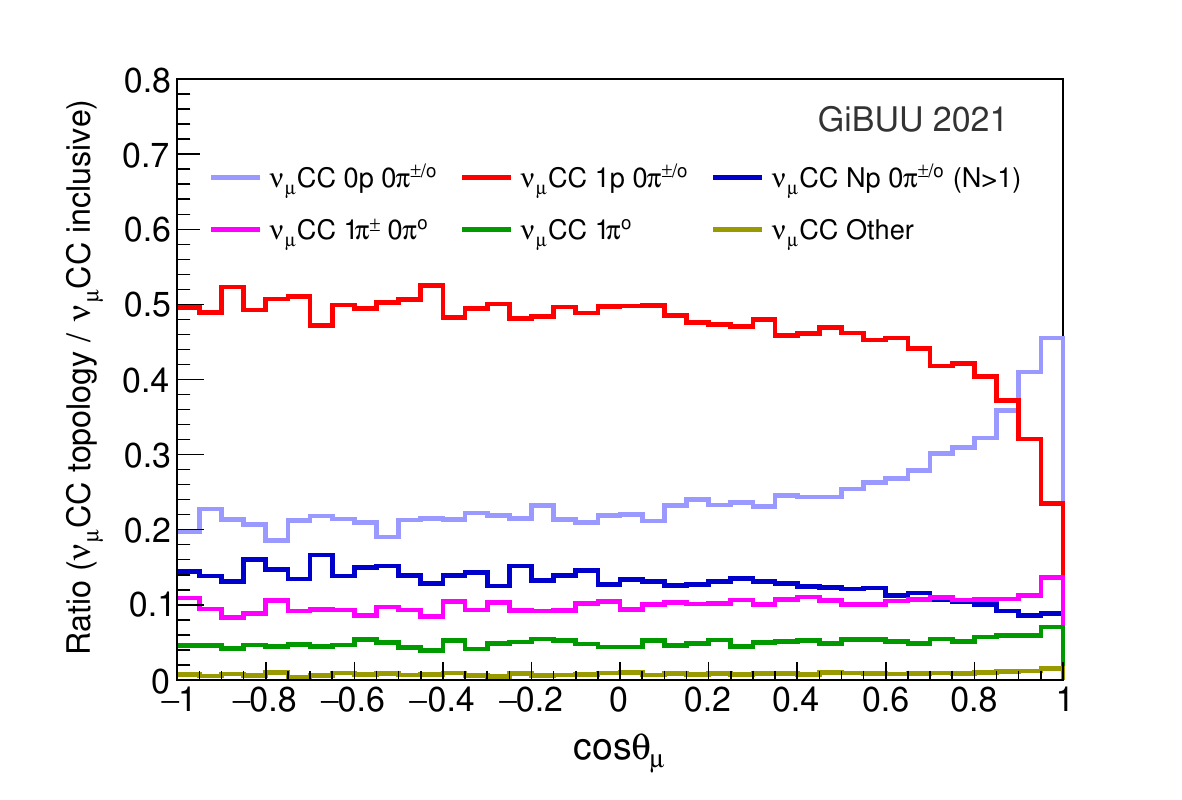}
    \includegraphics[width=0.45\textwidth]{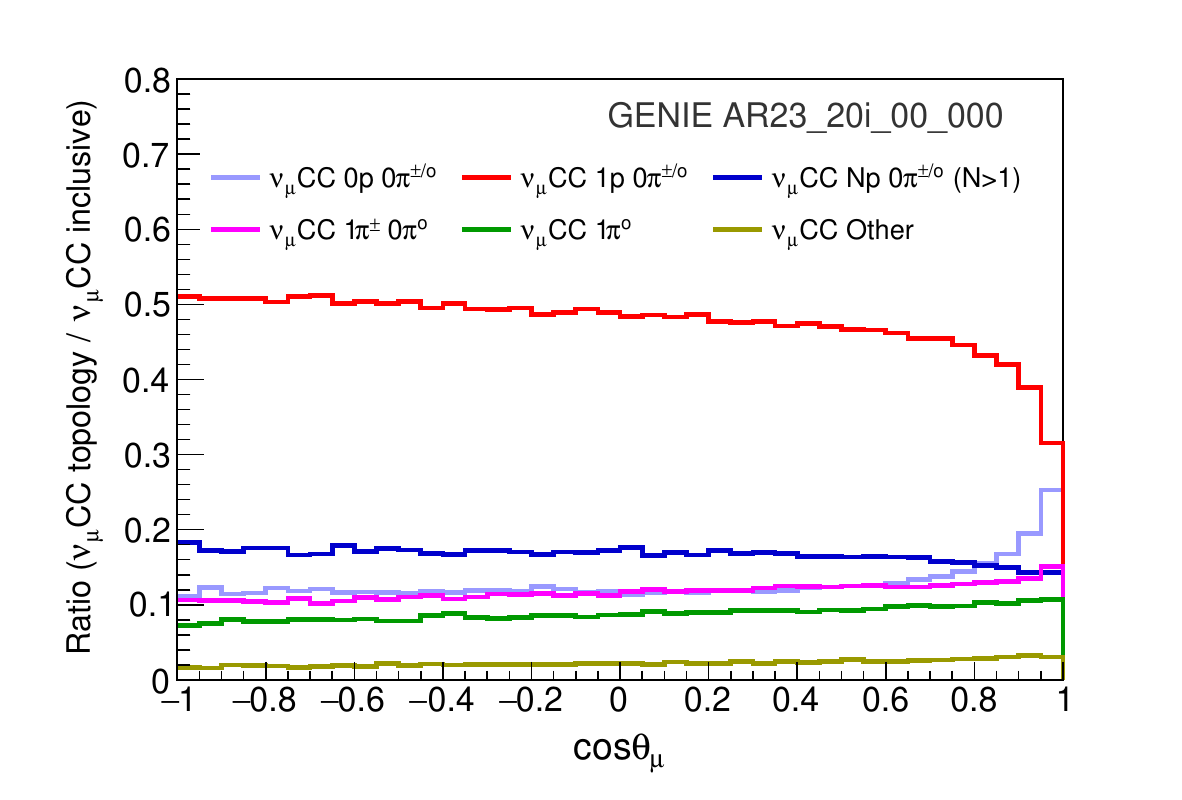}
    \includegraphics[width=0.45\textwidth]{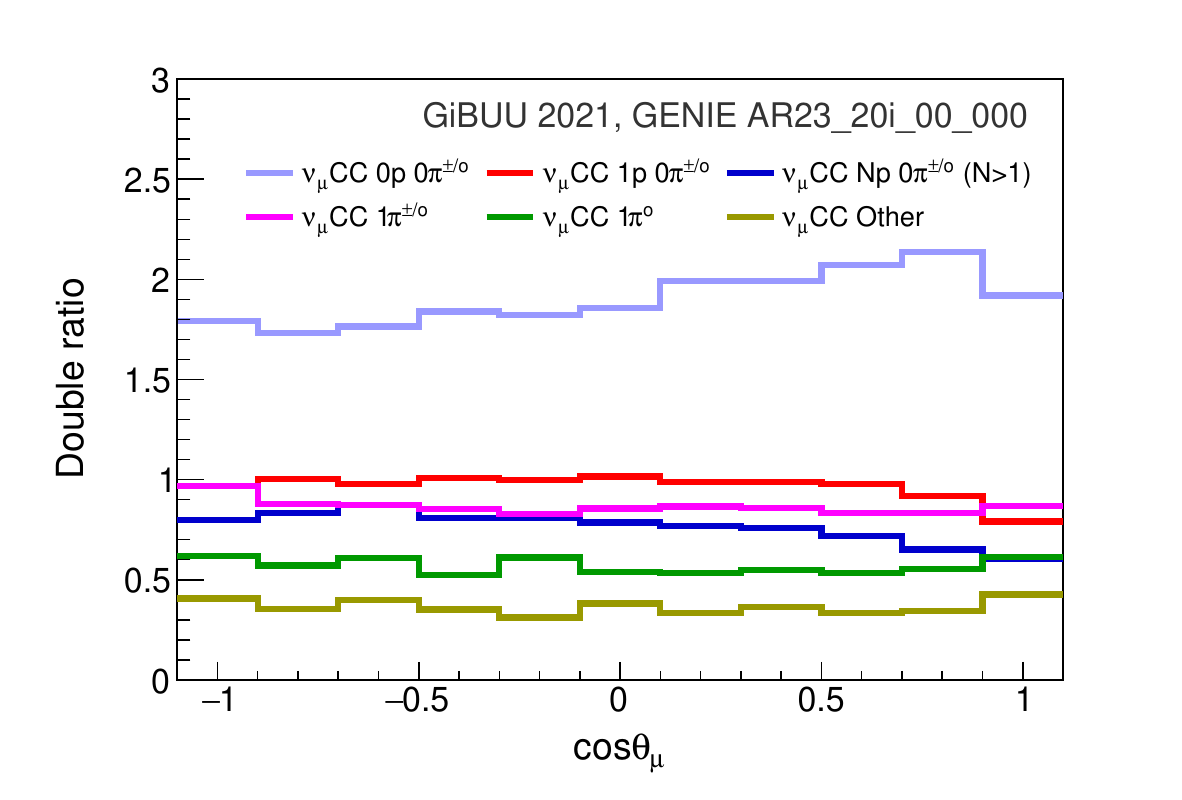}
    
    \caption{Ratios of each exclusive topology to the inclusive charged-current muon neutrino prediction as a function of muon angle, comparing GiBUU and GENIE in the LArBox geometry with the BNB flux. The top and middle plots show the ratios for GiBUU and GENIE, respectively. The bottom plot displays the double ratio, calculated as the ratio between the top and middle plots. Proton and pion kinetic energy thresholds are set at 50 MeV to define the topologies.}
    \label{fig:comptGenieGiBUU_cos}
\end{figure}

In this work, we performed studies using LArBox to compare GiBUU and GENIE predictions for the momentum (Fig.~\ref{fig:comptGenieGiBUU_mom}) and the cosine of the angle (Fig.~\ref{fig:comptGenieGiBUU_cos}) of outgoing muons in $\nu_{\mu} CC$ interactions. These figures highlight the nuanced differences between the models when generating events with the BNB neutrino flux at the LArBox detector location and geometry. 

We calculated the ratio of different neutrino interaction topologies relative to the inclusive $\nu_{\mu} CC$ events. These topologies include $\nu_{\mu} CC0P0\pi^{\pm /0}$ (no protons or pions are observed in the final state above an energy threshold, defined as 50 MeV in kinetic energy for protons and pions), $\nu_{\mu} CC1P0\pi^{\pm /0}$ (one proton and no pions are observed in the final state above the energy threshold), $\nu_{\mu} CCNP0\pi^{\pm / 0}$ (more than one proton and no pions are observed in the final state above the energy threshold), $\nu_{\mu} CC1\pi^{\pm} 0\pi^{0}$ (one charged pion, no neutral pions, and any number of protons are observed in the final state above the energy threshold), $\nu_{\mu} CC1\pi^{0}$ (one neutral pion, any number of charged pions, and any number of protons are observed in the final state above the energy threshold), $\nu_{\mu} CCOther$ (remaining charged current events not included in the previous exclusive topologies). The chosen particle energy thresholds  are realistic if considering LArTPC detectors and large enough to reduce the impact due to the particle identification efficiency at lower energies. 

Figure~\ref{fig:comptGenieGiBUU_mom} shows the ratio between a given neutrino-argon topology and the inclusive sample with respect to the muon momentum for GiBUU at the top and GENIE at the middle. The most noticeable difference is that the contributions of each topology are clearly different in each model. In GiBUU, the dominant contributions are $\nu_{\mu} CC1P0\pi^{\pm /0}$ and $\nu_{\mu} CC0P0\pi^{\pm /0}$, while in GENIE, the main contribution is $\nu_{\mu} CC1P0\pi^{\pm /0}$, with clear differences in the shape of the distributions across topologies between the models. The third plot in Figure~\ref{fig:comptGenieGiBUU_mom} is the ratio of the fractional contributions of each topology to the $\nu_{\mu} CC$ between GiBUU and GENIE to offer more details on the fractional differences between these two models.

\begin{figure}[!htbp]
    \centering
    \includegraphics[width=0.45\textwidth]{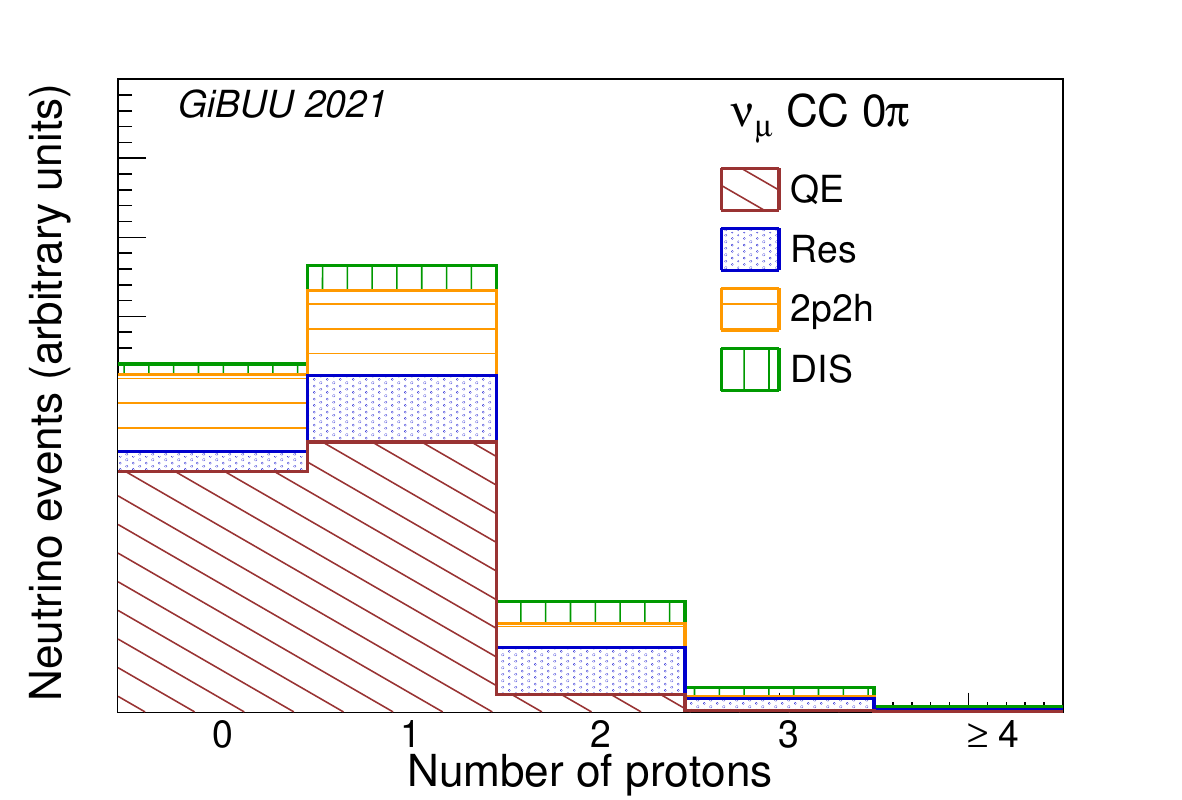}
    \includegraphics[width=0.45\textwidth]{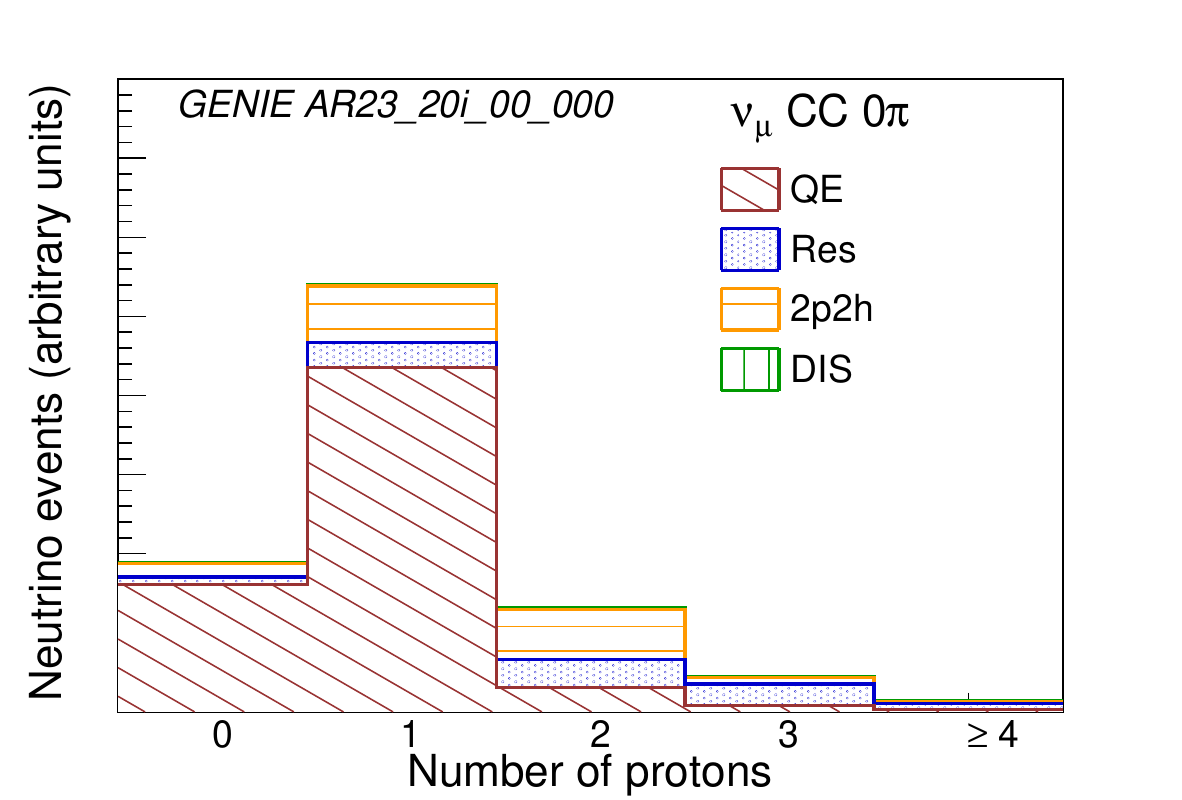}
    \caption{Proton multiplicity for the $\nu_{\mu}$ CC0$\pi^{\pm /0}$ topology in GiBUU (top plot) and GENIE (bottom plot). Proton and pion kinetic energy thresholds are set at 50 MeV in the topological definition.}
    \label{fig:protonmultiplicity}
\end{figure}

\begin{figure}[!htbp]
    \centering
    \includegraphics[width=0.45\textwidth]{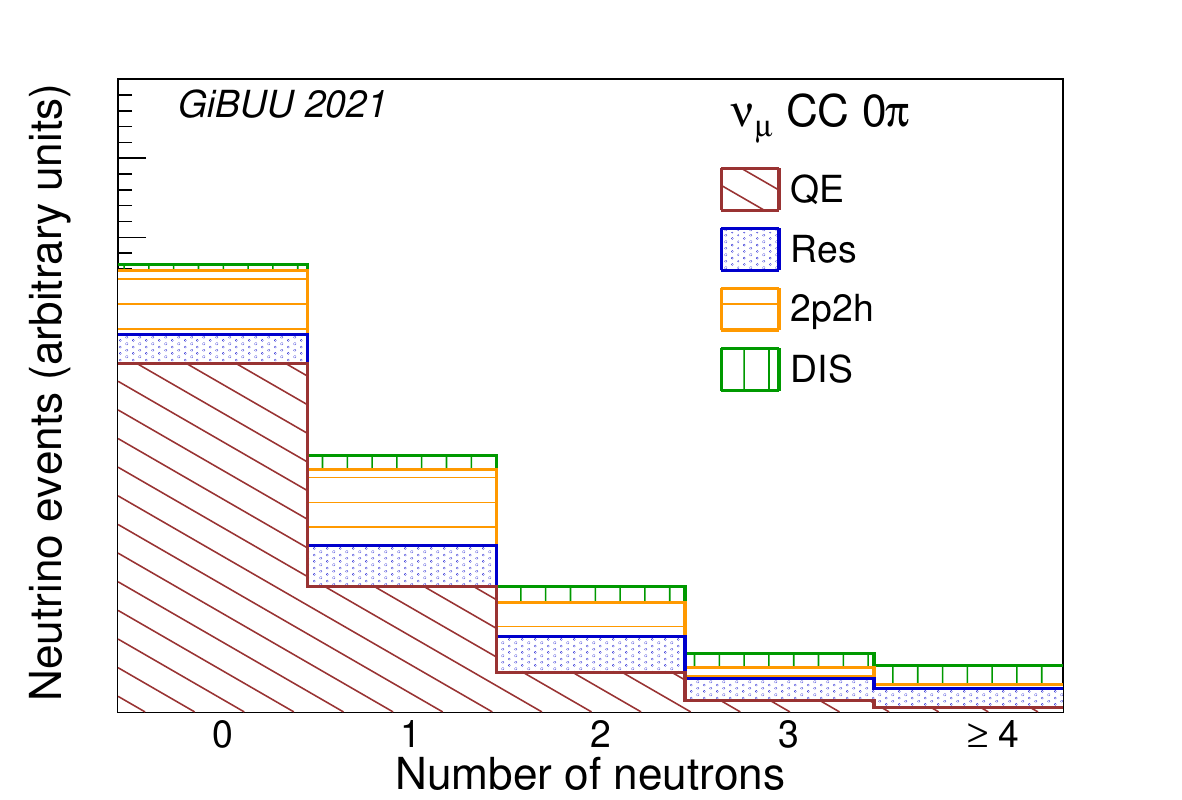}
    \includegraphics[width=0.45\textwidth]{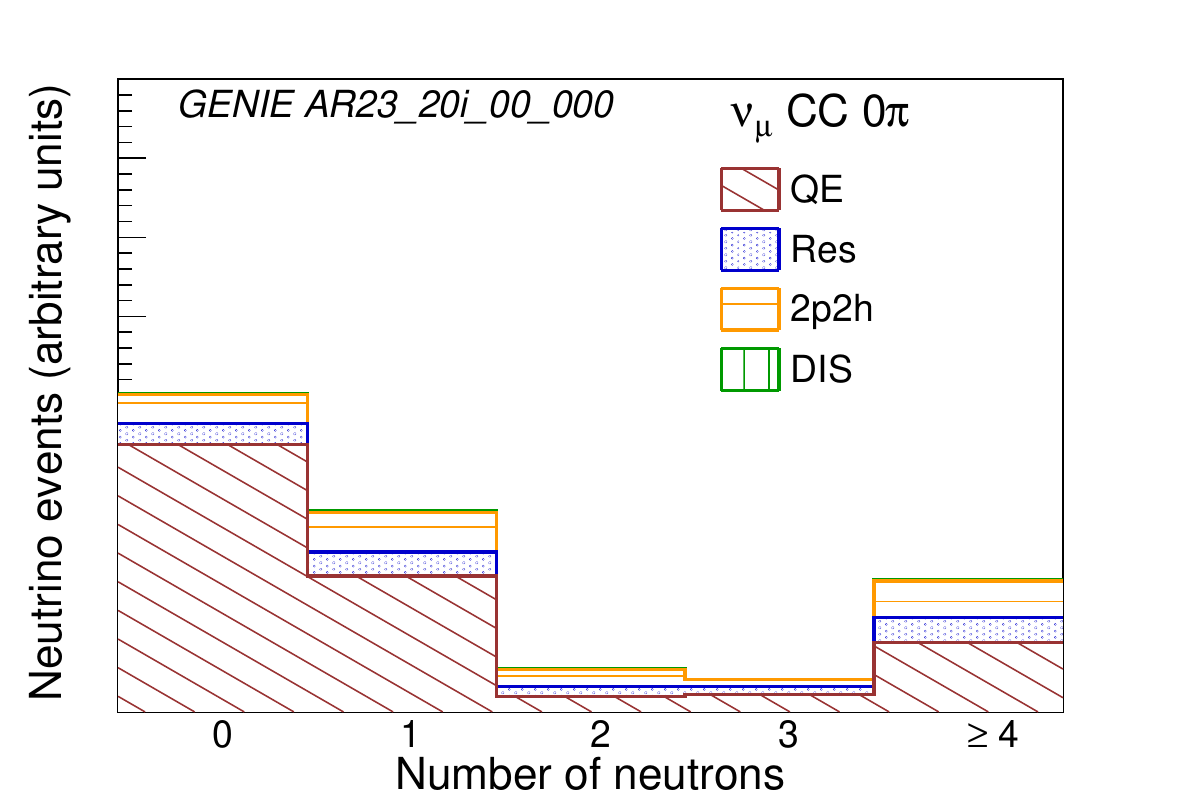}
    \caption{Neutron multiplicity for the $\nu_{\mu}$ CC0$\pi^{\pm /0}$ topology in GiBUU (top plot) and GENIE (bottom plot). Pion kinetic energy thresholds are set at 50 MeV in the topological definition.}
    \label{fig:neutronmultiplicity}
\end{figure}

By investigating different particle energy thresholds and particle multiplicities, we observed that GENIE produces a higher proton multiplicity than GiBUU. For example, as shown in Figure~\ref{fig:protonmultiplicity}, GENIE has a lower fraction of events with no protons below the energy threshold in the $\nu_{\mu} CC0\pi^{\pm /0}$ topology. Additionally, significant differences are observed in the descriptions of the models of neutron multiplicities, as illustrated in Figure~\ref{fig:neutronmultiplicity}, with large discrepancies between the generators. We also note distinct neutrino interaction compositions in each model, particularly in MEC and quasi-elastic production. 

One potential cause of these discrepancies is that GENIE does not include Pauli blocking in its treatment of FSI, resulting in many protons that should be absorbed in the nucleus being incorrectly retained. Furthermore, the treatment of FSI differs substantially between the two generators. These differences, along with other variations in nuclear medium modeling, likely explain the discrepancy in the $\nu_{\mu} CC0P0\pi^{\pm /0}$ topology between the models, which exhibited the largest difference in the double ratio shown in Figure~\ref{fig:comptGenieGiBUU_mom}.

Figure~\ref{fig:comptGenieGiBUU_cos} shows similar ratios, but with respect to the cosine of the muon angle relative to the neutrino direction. We observe a suppression of events for the most forward muons, which predominantly affect the $\nu_{\mu} CC1P0\pi^{\pm /0}$ topology in GiBUU and the $\nu_{\mu} CC0P0\pi^{\pm /0}$ topology in GENIE. These distributions are particularly important to understanding neutrino data, and the different trends of each Monte Carlo offer much information. The third plot shows the double ratio of these quantities, similar to Figure~\ref{fig:comptGenieGiBUU_mom}.

The different compositions by Monte Carlo tell us about the potential background predictions we can find in experimental data, which are very different depending on the model choice. These different composition predictions would also lead to important corrections in the neutrino energy reconstruction for an oscillation experiment and could produce to different outcomes.

In high-resolution detectors, such as LArTPC, the final observed particle content in the detector is critical for reconstructing neutrino energy. This approach, which aims to reduce model-dependent uncertainties compared to methods like CCQE-like approaches, faces significant challenges in identifying all hadrons emerging from the nucleus. A major hurdle is quantifying invisible neutrons and accurately modeling the final proton content and their kinematics. Correcting these observables, especially for nucleons, can introduce substantial uncertainties and biases in neutrino energy reconstruction.

\begin{figure}[!htbp]
    \centering
    \includegraphics[width=0.45\textwidth]{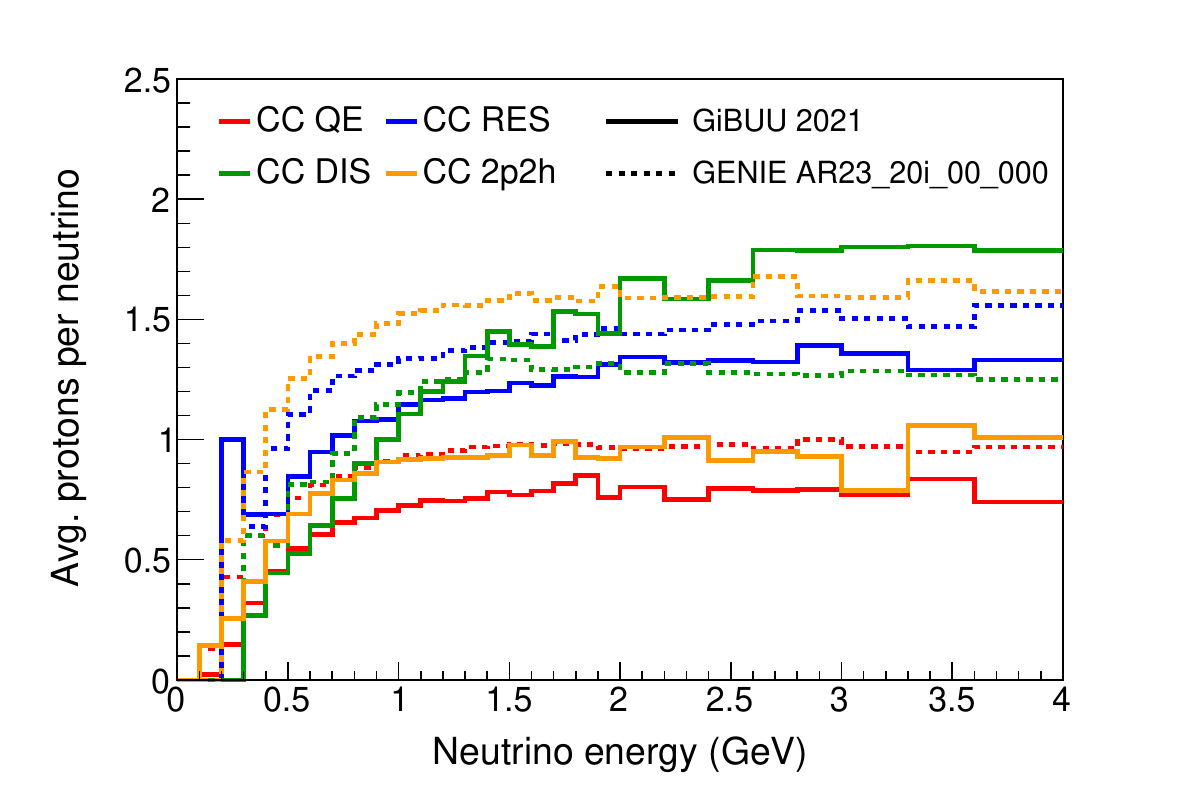}
    \caption{Average number of simulated protons above 50 MeV per neutrino at each neutrino energy, categorized by interaction type. Solid lines represent predictions from GiBUU, while dashed lines represent predictions from GENIE.}
    \label{fig:protonavgGenieGiBUU}
\end{figure}

Accurate modeling of the number of nucleons knocked out during neutrino interactions is essential, and access to diverse modeling approaches is crucial for tackling this challenge. In Figure~\ref{fig:protonavgGenieGiBUU}, we present the average number of protons generated per neutrino interaction across varying neutrino energies. Solid lines represent predictions from the GiBUU model, while dashed lines correspond to GENIE predictions. Generally, we observe that GENIE has a higher proton multiplicity than GiBUU, except for DIS above $\sim 1.4 GeV$. 

Notably, observing final-state protons assists in reconstructing the neutrino energy. For particles not observed—such as neutrons and low-energy particles below the detection threshold—model-based corrections are applied for oscillation measurements. This is one of the most model-dependent steps in oscillation data analysis, with the modeling of final-state nucleons being a substantial challenge for generators. This comparison highlights the nuanced differences between these models, which will directly impact neutrino energy reconstruction in experiments like SBN and DUNE.

The observed variations emphasize the significance of employing multiple models, such as GiBUU, to interpret neutrino scattering data accurately. These differences prompt further investigation into specific aspects of neutrino-nucleus interactions, refining our understanding and mitigating model-dependent uncertainties.

Validation against experimental data, including charged-current and neutral-current interactions, will demonstrate the model's capacity to faithfully reproduce a broad range of observables. Our study reinforces the importance of model diversity in refining the interpretation of neutrino scattering data, which is essential for pushing the boundaries of our understanding in this intricate field. The work developed is currently used by the SBND and ICARUS collaborations, and the framework will soon be integrated into LArSoft, enhancing accessibility across experiments. It can also be easily integrated into other experiments that do not use LArSoft, as long as they use GENIE for their event generator. For experiments that do not use GENIE, integration remains feasible by employing their event generator's flux and geometry drivers.

\section{Conclusion}
\label{sec:Conclusion}

In conclusion, implementing the GiBUU model as a Monte Carlo has yielded valuable insights into the complex dynamics of neutrino scattering processes. This Monte Carlo reproduces correctly the GiBUU model, and it is the first step in incorporating uncertainties to be propagated into experimental data analysis. Notably, a comparative study with GENIE reveals nuanced differences, emphasizing the necessity of employing multiple models to interpret neutrino scattering data comprehensively. The significant differences observed between GiBUU and GENIE, particularly in expected topological event samples and proton multiplicity (Figure~\ref{fig:comptGenieGiBUU_mom},~\ref{fig:protonmultiplicity}, and ~\ref{fig:comptGenieGiBUU_cos}), highlight the importance of understanding these models to better assess uncertainties  when reconstructing neutrino energy, especially for oscillation experiments. GiBUU offers a different approach to other generators with a unified nucleus description, bringing new opportunities to better explore our data and develop model uncertainties. The same methodology we used to incorporate GiBUU as an unweighted event generator could be used for other cross-section models not yet available as event generators, such as some Beyond the Standard Model theory simulations. Our simulation study employed the described technique to simulate GiBUU events within the LArBox geometry, utilizing the BNB flux.

While acknowledging tensions in specific datasets, GiBUU's overall robust performance establishes it as a reliable tool in the neutrino physics toolkit. The presented results underscore the importance of scrutinizing and refining model predictions, essential for advancing our comprehension of neutrino-nucleus interactions. 

\section{Acknowledgments}
\label{sec:Acknowledgements}

We thank Prof. Ulrich Mosel and Prof. Kai Gallmeister from Justus-Liebig-Universitat Giessen, Germany, for their indispensable guidance. Their expertise significantly enhanced our study of neutrino interactions.

\FloatBarrier
\bibliography{main}
\bibliographystyle{apsrev4-1}

\end{document}